\documentclass[12pt,twoside]{article}
\usepackage{amsmath}
\newcommand{\rd}[1]{\mathop{\mathrm{d}#1}}

\newcommand{\fract}[2]{{\textstyle\frac{#1}{#2}}}
\newcommand{\grad}{\vec\nabla}

\newcommand{\nA}{non-Abelian}
\newcommand{\Ab}{Abelian}
\newcommand{\CS}{Chern-Simons}
\newcommand{\CSt}{Chern-Simons term}
\newcommand{\CPt}{Chern-Pontryagin}
\newcommand{\Cpr}{Clebsch pa\-ra\-me\-ter\-iza\-tion}

\newcommand{\pr}{para\-me\-ter\-iza\-tion}
\newcommand{\prd}{para\-me\-ter\-ized}

\newcommand{\mn}{{\mu\nu}}
\newcommand{\ab}{{\alpha\beta}}
\newcommand{\pp}[1]{\partial_{#1}}

\newcommand{\numeq}[2]{\begin{equation}
#2
\label{#1}
\end{equation}}

\newcommand{\refpap}[1]{(\ref{pap#1})}

\let\vec\boldsymbol
\let\eps\varepsilon
\let\epsilon\varepsilon
\let\phi\varphi

\def\today{\ifcase\day\or
  1st\or 2nd\or 3rd\or 4th\or 5th\or
  6th\or 7th\or 8th\or 9th\or 10th\or
  11th\or 12th\or 13th\or 14th\or 15th\or
  16th\or 17th\or 18th\or 19th\or 20th\or
  21st\or 22nd\or 23rd\or 24th\or 25th\or
  26th\or 27th\or 28th\or 29th\or 30th\or
  31st\fi~\ifcase\month\or
  January\or February\or March\or April\or May\or June\or
  July\or August\or September\or October\or November\or December\fi
  \space \number\year}

\def\Journal#1#2#3#4{{\em #1} {\bf #2}, #3 (#4)}
\def\add#1#2#3{{\bf #1}, #2 (#3)}
\def\Book#1#2#3#4{{\em #1}  (#2, #3 #4)}
\def\Bookeds#1#2#3#4#5{{\em #1}, #2  (#3, #4 #5)}

\def\NPB{Nucl. Phys.} 
\def\PLA{Phys. Lett.} 
\def\PLB{Phys. Lett.} 
\def\PRL{Phys. Rev. Lett.}
\def\PRD{Phys. Rev. D}
\def\PR{Phys. Rev.}

\def\AnnP{Ann. Phys.\ ({\em NY})}

\def\PNAS{Proc. Nat. Acad. Sci.}
\def\PVAS{Proc. V.A. Steklov Inst. Math.}

\begin{document}
 
\title{Collaborating with David Gross;\\
Descendants of the Chiral Anomaly}
\author{R. Jackiw\\
\small\it Center for Theoretical Physics\\ 
\small\it Massachusetts Institute of Technology\\ 
\small\it Cambridge, MA 02139-4307}
\date{\small MIT-CTP-3089\quad Typeset in \LaTeX\ by M.
Stock\\[2ex]
\large\itshape In celebration of David Gross's birthday}
\maketitle

\abstract{\noindent
I recall my collaboration with David Gross. A result about descendants of the chiral
anomaly is presented: Chern-Simons terms can be written as total derivatives.
\null}

\pagestyle{myheadings}
\markboth{\emph{R. Jackiw} --- Collaborating with David Gross;}{\small 
Descendants of the Chiral Anomaly}
\thispagestyle{empty}

\section{Collaborating with David}

I am pleased to be the first celebrant of David Gross's significant birthday,
because among his colleagues and contemporaries in physics, I have known David the
longest -- since the time we both left our graduate schools to become Junior Fellows
at Harvard in 1966.   There our situation was not easy.  Harvard physics was
experiencing a downturn -- physics prominence at an institution undergoes cyclic
variation, like the economy.  David was coming from Chew's S-matrix theory at
Berkeley, I from Wilson's field theory at Cornell. Both Chew and Wilson were
prophets, but neither was recognized then at Harvard, except perhaps by Schwinger,
who was channeling their two streams into his source theory, with the motto ``If you
can't join 'em, beat 'em.''  Boston's Joint Theory Seminar had lapsed; Weinberg had
not yet arrived to invigorate the department.  A few miles away, MIT was bustling --
the S-matrix was in vogue and string theory in its first incarnation was being
developed by Veneziano and Fubini. I recall that David tried to connect with that
activity, but I guess it proved to be just too wearisome to travel even the small
distance from Harvard to MIT. Therefore the two of us were thrown together onto
our own resources, and if we were to collaborate with anyone, it had to be with each
other.  

So we did, and I learned that David is smarter than I.  This is because our
collaboration could happen in only one of two ways:  either I would learn S-matrix
theory, or David would learn field theory. Now there was no way that I could learn
S-matrix theory at that time -- I still can't --  therefore David had to learn field
theory.  As all of us know, he learned it quickly, he learned it well, he contributed to
it enormously, and then he left it -- again returning to the S-matrix   in its
M-theory--evolved form. 

Our collaboration
produced seven papers, and here they are: 

\begin{enumerate}

\item \label{pap1}
``Derivation of the SU(3) x SU(3) space-time local current
commutators'', \Journal{\PR}{163}{1688}{1967} 

\item \label{pap2}
``Low energy theorem for graviton scattering'', \Journal{\PR}{166}{1287}{1968} 

\item \label{pap3}
``Fermion avatars of the Sugawara model'' (with S.~Coleman),
\Journal{\PR}{180}{1359}{1969} 

\item \label{pap4}
``Construction of covariant and gauge invariant T* products'',
\Journal{\NPB}{B14}{269}{1969} 

\item \label{pap5}
``Dimensions of currents and current commutators'' (with M.~Beg, J.~Bernstein, and
A. Sirlin),
\Journal{\PRL}{25}{1231}{1970} 

\item \label{pap6}
``Effects of anomalies on quasi-renormalizable theories'',
\Journal{\PRD}{6}{477}{1972} 

\item \label{pap7}
``Constraints on anomalies''  (with S.~Adler and
C.~Callan), \Journal{\PRD}{6}{2982}{1972} 

\end{enumerate}
Although Weinberg was not
physically present, his influence is evident in that all the papers deal in one way or
another with currents and current algebra, subjects emphasized by Weinberg, whom
David already knew at Berkeley.  But looking at these papers again, I now see that the
topics researched by us belong to the late, baroque period of current algebra, when
the formalism became encrusted by various commutator and current conservation
anomalies.  Thus papers \refpap1 and \refpap4 deal with commutator anomalies: 
Schwinger terms and seagull terms,  as does paper \refpap3 where we use the
Schwinger term in fermionic current commutators to fit a theory with Dirac fields into
the Sugawara form~\cite{ref1A} -- a construction that years later became popular and
was repeated many times. In paper
\refpap5 we use the Schwinger term to describe the total electroproduction cross
section, while paper~\refpap7 relies on the Adler-Bardeen anomaly
non-renormalization theorem~\cite{ref2A} to show that the Gell-Mann--Low~function
of quantum electrodynamics~\cite{ref3A} has no zeroes.  Our most famous paper --
{\sl\uppercase{Spires}} calls it a ``renowned paper'' --  concerns anomalies in the
standard model.  In fact, the collaboration was a long-distance one;  we had left
Harvard, David for Princeton and I for MIT.  The paper began as a separate,
independent effort.  I was finishing my manuscript while visiting the newly
constituted theory group at Fermilab, and Treiman, its temporary leader, informed
me that David was completing similar research.  Joining our efforts turned out to be a
great benefit for me, because  again I profited from David's positive approach:  I had
put forward the negative result that owing to the chiral anomaly the
Weinberg-Salam model was not renormalizable, contrary to 't~Hooft's and Veltman's
claims; while concurring, David put forward the principle of anomaly cancellation to
save renormalizability~\cite{ref4A}.  Thus our work provides one of two examples
that not only theoretical physicists  but also Nature makes use of the chiral anomaly
(the other example being neutral pion decay~\cite{ref5A}).

My favorite paper is our second, which is not ``renowned''; according to
{\sl\uppercase{Spires}} it is less than ``unknown'' -- it is unlisted. Here we are
recalling the famous Compton scattering low-energy theorems due to Thirring,
Gell-Mann, Goldberger, and Low~\cite{ref6A}, which are consequences of gauge
invariance, and are derived with the help of Ward identities. It seemed to us that
diffeomorphism invariance should entail similar low-energy theorems for graviton
scattering, but we did not know at that time how to construct the analogous Ward
identities~\cite{ref7A}.   David's S-matrix expertise again made progress possible.
It happened that just in those days Abarbanel and Goldberger rederived the
Compton scattering results using S-matrix dispersive methods~\cite{ref8A}. David
suggested that we adopt that approach for our graviton problem, and we succeeded
in doing the impossible: we derived an exact result for quantum gravity, even though
quantum gravity theory did not then exist, and perhaps still doesn't!

Even after our physical collaboration ended, we continued to work together
unconsciously, in spirit, because four years after our last joint paper, David (together
with Callan and Dashen~\cite{ref9A})  and I (together with Rebbi~\cite{ref10A})
produced identical analyses of the Yang-Mills vacuum and its angle.

Probably David would prefer hearing about new results, while I have been
reminiscing about old ones. So let me conclude with something that he, and all of you,
may find surprising.

\section{Descendants of the Chiral Anomaly}

By now we appreciate that anomalous divergences of chiral fermionic currents
involve topological \CPt\ densities. In four dimensions we have the \Ab\ and
\nA\ formulas:
\begin{align}
\mbox{anomaly}&= {}^*\!F^{\mn} F_\mn = 
\fract12 \eps^{\mn\ab} F_\mn F_\ab 
&\mbox{(\Ab)}\label{eq-1}\\
\mbox{anomaly}&= {}^*\!F^{\mn a} F_\mn^a  =  
\fract12 \eps^{\mn\ab} F_\mn^a   F_\ab^a    &\mbox{(\nA);}
\label{eq-2}\\
\intertext{in two dimensions there is the \Ab\ expression:}
\mbox{anomaly}&= {}^*\!F = \fract12 \eps_\mn F^\mn  &\mbox{(\Ab).}\label{eq-3}
\end{align}
Analogous formulas hold in all other even dimensions~\cite{ref6}.

These quantities are topologically interesting and their volume integrals are
topological invariants, measuring various topological characteristics of gauge fields.
Consequently one expects that all these quantities can be presented as total
derivatives, so that the volume integral becomes converted by Gauss's law into a
surface integral, sensitive only to long-distance, global properties of the gauge fields,
as befits a topological entity. This is indeed the case, but the possibility of expressing
the \CPt\ densities as total derivatives emerges only when the field strengths are
presented in terms of potentials. In four dimensions:
\begin{align}
F_\mn &= \pp\mu A_\nu - \pp \mu A_\nu\nonumber\\
\fract12 {}^*\!F^\mn F_\mn  &= \pp \mu \bigl(\eps^{\mu\alpha\beta\gamma}
A_\alpha \pp\beta A_\gamma \bigr) & \mbox{(\Ab)}\label{eq-4}\\[2ex]
F_\mn^a &= \pp\mu A_\nu^a - \pp \mu A_\mu^a + f^{abc}
A_\mu^b A_\nu^c\nonumber\\
\fract12 {}^*\!F^{\mn a} F_\mn^a  &= \pp\mu \eps^{\mu\alpha\beta\gamma}
\bigl( A_\alpha^a\pp\beta A_\gamma^a + \fract13 f^{abc} A_\alpha^a A_\beta^b
A_\gamma^c
\bigr)& \mbox{(\nA);} \label{eq-5}\\
\intertext{in two dimensions:}
F_\mn &= \pp\mu A_\nu -\pp\nu  A_\mu\nonumber\\
 {}^*\!F &= \pp\mu (\eps^\mn A_\nu) & \mbox{(\Ab).} \label{eq-6}
\end{align}

The quantities whose divergences give the even-dimensional \CPt\ densities  are
called \CS\ terms. By suppressing one dimension, they become naturally defined on
an odd-dimensional manifold, in one lower  dimension,  and we are thus led to
consider the
\CS\ terms in their own right~\cite{ref7}. In three dimensions:
\begin{align}
\mathrm{CS}(A) &= \eps^{ijk} A_i \pp j A_k & \text{(Abelian)}\label{eq7}\\
\mathrm{CS}(A) &= \eps^{ijk} \bigl(A_i^a \pp j A_k^a +\fract13 f^{abc} A_i^a A_j^b
A_k^c\bigr)
 & \text{(non-Abelian);}\label{eq8}\\
\intertext{in one dimension:}
\mathrm{CS}(A) &= A_1 &\mbox{(\Ab).}\label{eq-9}
\end{align}

The (3 and 1)-dimensional volume integrals of these quantities are again topological
invariants, with interesting physical information. The three-dimensional integral in
the \Ab\ case -- the case of electrodynamics -- is called the magnetic helicity: 
$\int \rd{^3 r} \vec A\cdot \grad
\times \vec A = 
\int \rd{^3 r} \vec A\cdot \vec B$ and measures the linkage of magnetic flux lines.
An analogous quantity arises in fluid mechanics  with the local fluid velocity $\vec v$
replacing
$\vec A$  and  vorticity $\vec\omega = \grad\times\vec v$ replacing~$\vec B$.
Then the integral $\int \rd{^3 r} \vec v\cdot \grad\times \vec v = \int \rd{^3 r} \vec
v\cdot \vec \omega$ is called kinetic vorticity~\cite{ref8}. A nonvanishing kinetic
vorticity presents an obstacle to a canonical formulation for the Euler equations of
fluid mechanics~\cite{ref10}. Yet another property of the volume integral of the \CSt\
is that when its vector potential evaluated on a pure gauge configuration, the integral 
measures the windings of the gauge function~\cite{ref6,ref7}.

I shall not review here the many uses that arose after my collaborators and I
introduced  the Abelian and
\nA\ 
\CSt s~\cite{ref7}. The applications range from the
mathematical characterization of knots to the physical description of electrons in the
quantum Hall effect~\cite{ref9},  vivid evidence for the deep significance of the \CS\
structure and of its antecedent, the chiral anomaly.

Instead I pose the following question: Can one write the \CSt\ as a total derivative, so
that (as befits a topological quantity) the spatial volume integral becomes a surface
integral. An argument that this should be possible is the following: The
\CSt s that we have considered are a 3-form on 3-space and a 1-form on the line;
hence they are maximal forms, and their exterior derivatives vanish because there
are no 4-forms on 3-space nor 2-forms on a line. This establishes that the \CSt s are
closed, so one can expect that they are also exact, at least locally; that is, they can be
written as a total derivative.  But on their respective manifolds,  such derivative 
representations for the
\CSt s require expressing the potentials in terms of ``pre-potentials'', since the 
above \CS\ formulas show no evidence of derivative structure.
[Recall that the total derivative formulas  for
the axial anomaly also require using potentials to express~$F$.]

There is a physical, practical reason for wanting the three-dimensional, Abelian \CSt\ 
to be a total derivative. It is known in fluid mechanics that there exists an
obstruction to constructing a Lagrangian for Euler's fluid equations, and this
obstruction is just the kinetic helicity  \hbox{$\int \rd{^3 r} \vec v\cdot
\vec\omega$}, that is, the volume integral of the Abelian \CSt, constructed from the
velocity 3-vector~$\vec v$. This obstruction is removed when the integrand is a total
derivative, because then the kinetic helicity volume integral is converted to a surface
integral by Gauss' theorem. When the integral obtains contributions only from a
surface,  the obstruction disappears from the 3-volume, where the fluid equation
acts~\cite{ref10}. 

It is easy to show that the  Abelian \CSt\ can be presented as a total derivative. In
one dimension the result is trivial: any function can be written as a derivative of
another function:
$$
A_1 = \pp1 \theta \ .
$$
In three dimensions, we use the \Cpr\ for  a 3-vector~\cite{ref11}
\numeq{eq10}{
\vec A = \grad\theta + \alpha\grad\beta\ .
}
This nineteenth-century \pr\ of the a 3-vector $\vec A$ in terms of the
pre-potentials ($\theta$, $\alpha$, $\beta$) is an alternative to the usual
transverse/longitudinal
\pr. In modern language it is a statement of Darboux's
theorem that the 1-form $A_i \rd{x^i}$ can be written as $\rd \theta + \alpha
\rd\beta$~\cite{ref12}. With this
\pr\ for $\vec A$, one sees that the Abelian \CSt\ is indeed a total derivative:
\begin{align}
\mathrm{CS}(A) &= \eps^{ijk} A_i \pp j A_k\label{eq11}\\
&=  \eps^{ijk} \pp i \theta\pp j \alpha \pp k \beta\nonumber\\
&= \pp i \bigl( \eps^{ijk} \theta\pp j \alpha \pp k \beta\bigr)\ .\nonumber
\end{align}

When the \Cpr\ is employed for $\vec v$ in the fluid dynamical context, the
obstruction to a canonical formulation is removed, and the situation is analogous to
the force law in electrodynamics. While the Lorentz equation is written in terms of
field strengths, a Lagrangian formulation needs potentials from which the field
strengths are reconstructed. Similarly, Euler's equation involves the velocity
vector~$\vec v$, but in a Lagrangian for this equation the velocity must be
parameterized in terms of the  prepotentials
$\theta$, $\alpha$, and~$\beta$. 

When the \Cpr\ is employed for the electromagnetic vector potential $\vec A$,
magnetic helicity acquires an appealing form:
\numeq{eq12}{
\int \rd{^3 r} \vec A\cdot (\grad \times \vec A) = 
\int \rd{^3 r} \grad \cdot (\theta\vec B) = \int \rd{\vec S} \cdot\, \theta\vec B\ . 
}
The magnetic helicity is the flux of the magnetic field through the surface bounding
the volume, with  $\theta$ acting as a modulating factor.

In a natural generalization of the above, one asks whether a \nA\ vector potential
can also be \prd\ in such a way that the \nA\ \CSt~\eqref{eq8} becomes a total
derivative. We have answered this question affirmatively and we have found
appropriate prepotentials that do the job~\cite{ref10,ref13,ref14}, but the details of
the construction are too technical to be presented here. We hope that our \nA\
generalization of the \Cpr\ will be as interesting and useful as the Abelian one,
perhaps for a \nA\ version of fluid mechanics.

\vspace*{-\bigskipamount}

\end{document}